\documentclass[conference]{IEEEtran}
\IEEEoverridecommandlockouts
\usepackage{cite}
\usepackage{amsmath,amssymb,amsfonts}
\usepackage{graphicx}
\usepackage{textcomp}
\usepackage{xcolor}
\usepackage{multirow}
\usepackage{mathtools,xparse}
\usepackage[dvips]{epsfig}
\usepackage{adjustbox}
\usepackage{subcaption}

\usepackage{array}
\usepackage{makecell}
\usepackage{multicol}

\usepackage{mathabx}
\usepackage{caption}
\usepackage{algorithm}
\usepackage{algpseudocode}

\usepackage{tabularx}
\usepackage{float}

\PassOptionsToPackage{hyphens}{url}\usepackage{hyperref}

\def\BibTeX{{\rm B\kern-.05em{\sc i\kern-.025em b}\kern-.08em
    T\kern-.1667em\lower.7ex\hbox{E}\kern-.125emX}}
\begin{document}

\title{ExcitNet Vocoder: A Neural Excitation Model for Parametric Speech Synthesis Systems
}

\author{\IEEEauthorblockN{Eunwoo Song$^{1}$, Kyungguen Byun$^{2}$ and Hong-Goo Kang$^2$}
\IEEEauthorblockA{\textit{$^1$NAVER Corp., Seongnam, Korea}\\
\textit{$^2$Department of Electrical and Electronic Engineering, Yonsei University, Seoul, Korea}\\
eunwoo.song@navercorp.com}
}

\maketitle
\fontsize{9.4}{10.5}\selectfont
	\begin{abstract}
    This paper proposes a WaveNet-based neural excitation model (ExcitNet) for statistical parametric speech synthesis systems.
    Conventional WaveNet-based neural vocoding systems significantly improve the perceptual quality of synthesized speech by statistically generating a time sequence of speech waveforms through an auto-regressive framework.
    However, they often suffer from noisy outputs because of the difficulties in capturing the complicated time-varying nature of speech signals.
    To improve modeling efficiency, the proposed ExcitNet vocoder employs an adaptive inverse filter to decouple spectral components from the speech signal.
    The residual component, i.e. excitation signal, is then trained and generated within the WaveNet framework.
    In this way, the quality of the synthesized speech signal can be further improved since the spectral component is well represented by a deep learning framework and, moreover, the residual component is efficiently generated by the WaveNet framework. 
    Experimental results show that the proposed ExcitNet vocoder, trained both speaker-dependently and speaker-independently, outperforms traditional linear prediction vocoders and similarly configured conventional WaveNet vocoders.
	\end{abstract}
	

	\section{Introduction}\label{sec:intro}

    Statistical parametric speech synthesis (SPSS) systems are popularly used for various applications, and much research has been performed to analyze the relationship between the accuracy of vocoding techniques and the quality of synthesized speech \cite{agiomyrgiannakis2015vocaine, raitio2014voice, hu2015fusion, song2017effective}.
    In the typical source-filter theory of speech production \cite{quatieri2006discrete}, the residual signal, i.e. source, is obtained by passing the speech signal through a linear prediction (LP) filter that decouples the spectral formant structure.
    To reduce the amount of information, the residual signal is approximated by various types of excitation model such as pulse or noise (PoN) \cite{yoshimuray1999simultaneous}, band aperiodicity (BAP) \cite{kawahara1997speech, morise2016world}, glottal excitation \cite{raitio2011hmm,airaksinen2017glottal}, and time-frequency trajectory excitation (TFTE) models \cite{song2015improved}.
    As parametric vocoding techniques have become more sophisticated, so the quality of synthesized speech has improved. 
    
    Recently, WaveNet-based waveform generation systems have attracted great attention in the speech signal processing community thanks to their high performance and ease of application \cite{van2016wavenet}.
	In this type of system, the time-domain speech signal is represented as a discrete symbol sequence and its probability distribution is autoregressively modeled by stacked convolutional layers.
	By appropriately conditioning the acoustic parameters with input features, these systems have also been successfully adopted into neural vocoder structures \cite{tamamori2017speaker, hayashi2017investigation, hu2017ustc, shen2017natural, tachibana2018investigation, adiga2018use}.
	By directly generating the time sequence of speech signals without utilizing parametric approximation, WaveNet-based systems provide superior perceptual quality to traditional linear predictive coding (LPC) vocoders \cite{tamamori2017speaker}.

	However, the speech signals generated by a WaveNet vocoder often suffer from noisy outputs because of the prediction errors caused by adopting convolutional neural network (CNN) models. 
	Due to difficulties in capturing the dynamic nature of speech signals, spectral distortion can increase, especially in the high frequency region.
	Using properties of the human auditory system \cite{schroeder1979optimizing},
	Tachibana et. al. introduce a perceptual noise-shaping filter as a pre-processing stage in the WaveNet training process \cite{tachibana2018investigation}. 
	Although this approach improves the perceived quality of the generated speech, its modeling accuracy is relatively low in unvoiced and transition regions.
	The reason for this can be found from the time-invariant limitation of the noise-shaping filter which is not appropriate for regions where phonetic information varies significantly.

    To alleviate the aforementioned problem, we propose \textit{ExcitNet}; a WaveNet-based neural excitation model for speech synthesis systems.
    The proposed system takes advantage of the merits of both the LPC vocoder and the WaveNet structure.
    In the analysis step, the LP-based adaptive predictor is used to decouple the spectral formant structure from the input speech signal \cite{atal1979predictive}.
    The probability distribution of its residual signal, i.e., the excitation, is then modeled by the WaveNet framework.
    As the spectral structure represented by LP coefficients, or by their equivalents such as line spectral frequencies (LSFs), is changing relatively slowly, it is easy to model with a simple deep learning framework \cite{song2017effective}.
    In addition, because the variation of the excitation signal is only constrained by vocal cord movement, the WaveNet training process becomes much simpler.
    Furthermore, we significantly improve the WaveNet modeling accuracy by adopting the improved time-frequency trajectory excitation (ITFTE) parameters as conditional features that effectively represent the degree of periodicity in the excitation signal \cite{song2015deep, song2016improved, song2017perceptual}.

    In the speech synthesis step, an acoustic model designed using a conventional deep learning-based SPSS system first generates acoustic parameters from the given input text \cite{song2017effective}.
    Those parameters are used to compose input conditional features and the WaveNet then generates the corresponding time sequence of the excitation signal.
    Finally, the speech signal is reconstructed by passing the generated excitation signal through the LP synthesis filter.

    We investigate the effectiveness of the proposed ExcitNet vocoder not only in the analysis/synthesis but also in the SPSS frameworks.
    Our experiments verify the performance of various types of neural vocoders including a plain WaveNet \cite{tamamori2017speaker, hayashi2017investigation}, a noise shaping filter-based WaveNet \cite{tachibana2018investigation}, and a proposed ExcitNet trained both speaker-dependently and speaker-independently.
    The synthesis quality of each system is investigated by varying the amount of the training data set, of which results could be usefully referred to when designing similarly configured WaveNet-based neural vocoding frameworks.  
    Regarding the vocoder itself, in a perceptual listening test, the proposed system shows superiority over the our best prior parametric ITFTE vocoder with the same SPSS model structure.

\section{WaveNet Vocoder}
\label{sec:ch2}
    The basic WaveNet is an autoregressive network which generates a probability distribution of waveforms from a fixed number of past samples \cite{van2016wavenet}.
    The joint probability of the waveform is factorized by a product of conditional probabilities as follows:
	\begin{equation}\label{eq:wn}
	p(\mathbf{x})=\prod\limits_{t=1}^{T}{p({{x}_{t}}|{{x}_{1}},...,{{x}_{t-1}})},
	\end{equation}
    where $\mathbf{x}=\{{{x}_{1}},...,{{x}_{T}}\}$ denotes discrete waveform symbols compressed via $\mu$-law companding transformation.
    Given an additional input $\mathbf{h}$, defined as the auxiliary features, the WaveNet is able to model the conditional distribution $p(\mathbf{x}|\mathbf{h})$ of the waveform \cite{van2016wavenet}.
    By conditioning the model on other input variables, the output can be guided to produce waveforms with required characteristics.
    Typically, the original WaveNet uses linguistic features, fundamental frequency (F0), and/or speaker codes for the auxiliary condition \cite{van2016wavenet}.
    More recent WaveNet vocoders utilize acoustic parameters directly extracted from the speech such as mel-filterbank energy, mel-generalized cepstrum, BAP, and F0 \cite{tamamori2017speaker, hayashi2017investigation, hu2017ustc, shen2017natural, tachibana2018investigation, adiga2018use}.
    This enables the system to automatically learn the relationship between acoustic features and speech samples which results in superior perceptual quality over traditional LPC vocoders \cite{tamamori2017speaker, wang2018comparison}.

    \begin{figure}[!t]
    \centerline{\epsfig{figure=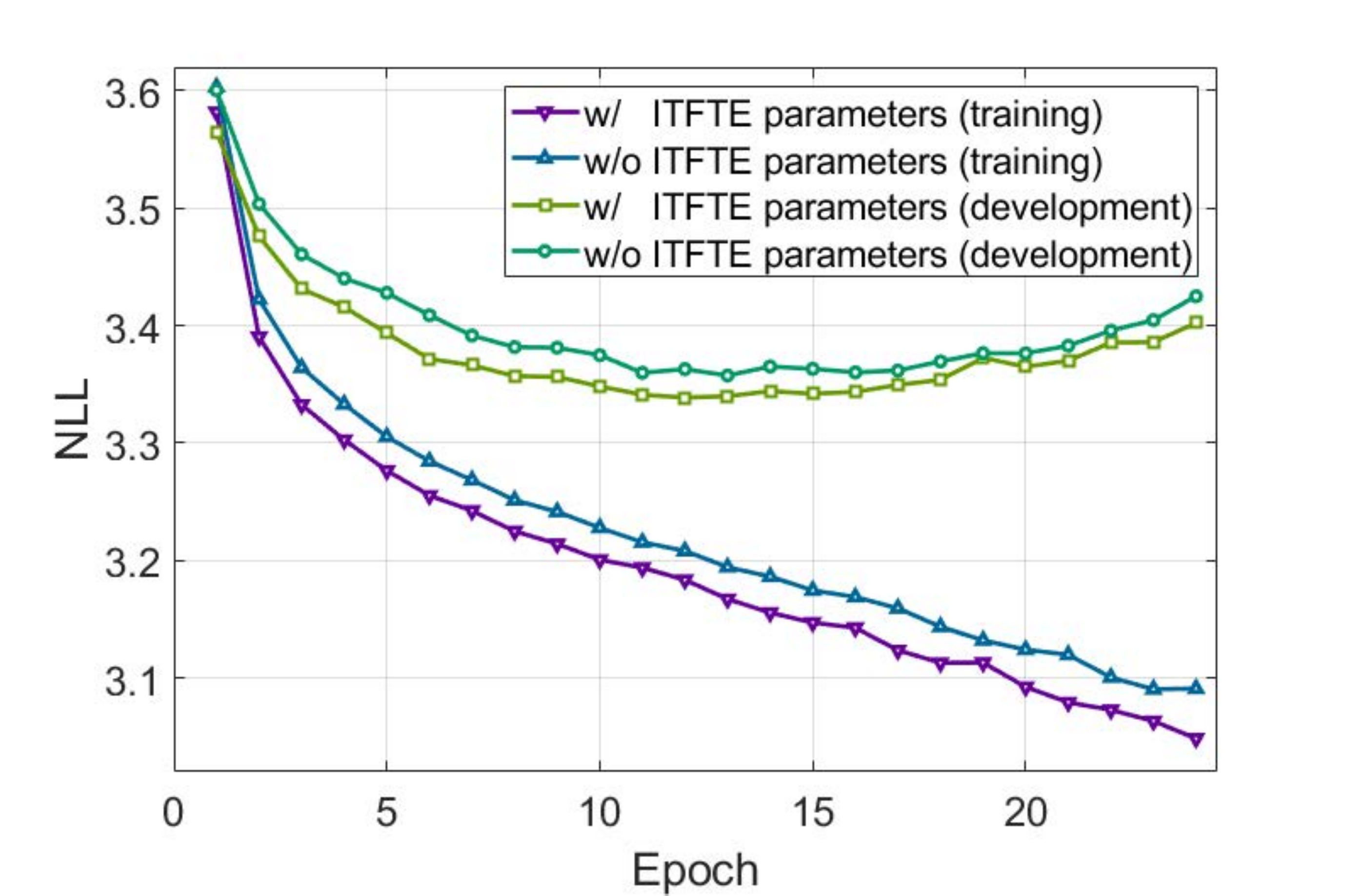,width=70mm}}
    \caption{Negative log-likelihood (NLL) obtained during the training process with (w/) and without (w/o) ITFTE parameters.}
    \label{fig:nll}
    \end{figure}

\section{ExcitNet vocoder}
\label{sec:ch3}
    Even though previous studies have indicated the technical possibility of introducing WaveNet-based vocoding, the systems often suffer from noisy outputs because of prediction errors in the WaveNet models \cite{tachibana2018investigation}.
    Since it is still challenging for CNNs to fully capture the dynamic nature of speech signals, more sophisticated system architecture that can effectively remove the redundant structures needs to be designed.
    
    In this research, we propose the ExcitNet vocoder; a neural excitation model for speech synthesis systems.
    In the proposed method, the redundant formant structure of the speech signal is removed using an LP analysis filter and the distribution of its residual signal, i.e., the excitation, is then modeled by a WaveNet framework.

\subsection{Auxiliary Features Employing ITFTE Parameters}
\label{ssec:ch3-1}  
    Similar to the conventional WaveNet vocoders, the input auxiliary features are composed of the spectral parameters, i.e., LSF, F0, v\texttt{/}uv, and gain.
    To further improve training efficiency, we also adopt ITFTE parameters \cite{song2015improved}.
    Note that the TFTE represents the spectral shape of excitation along the frequency axis and the evolution of this shape along the time axis.
    To obtain the harmonic excitation spectrum, i.e., a slowly evolving waveform (SEW), each frequency component of the TFTE is low-pass filtered along the time-domain axis.
    Beyond the cut-off frequency, the remaining noise spectrum, i.e., a rapidly evolving waveform (REW), is obtained by subtracting the SEW from the TFTE. 
    
    Employing the SEW and REW enables to effectively represent a periodicity distribution of the excitation \cite{song2015deep, song2016improved, song2017perceptual}.
    Therefore, adding these parameters to the auxiliary features helps improve the modeling accuracy of the WaveNet.
    Fig.~\ref{fig:nll} shows the negative log-likelihood obtained during the training process, of which result confirms that composing auxiliary features with ITFTE parameters enables to reduce both training and development errors as compared to the process without ITFTE parameters.
      
   	\begin{figure}[!t]
   	\begin{minipage}[b]{1.0\linewidth}
   		\centerline{\epsfig{figure=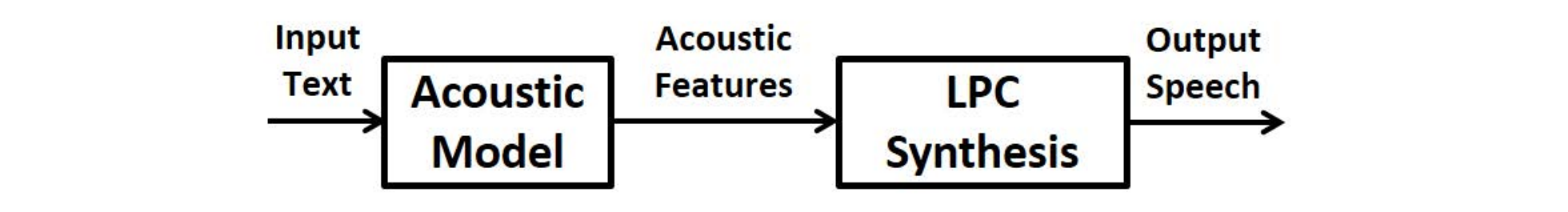,width=80mm}}
   		\centerline{(a)}  \medskip
   	\end{minipage}
   	\begin{minipage}[b]{1.0\linewidth}
   		\centerline{\epsfig{figure=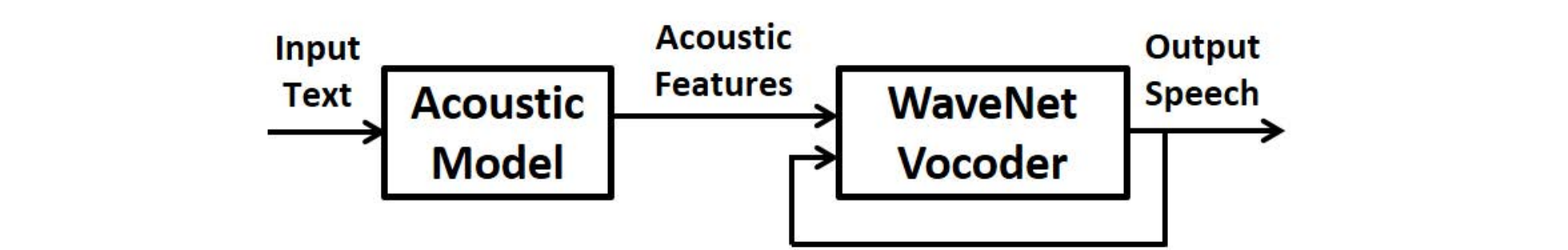,width=80mm}}
   		\centerline{(b)}  \medskip
   	\end{minipage}
   	\begin{minipage}[b]{1.0\linewidth}
   		\centerline{\epsfig{figure=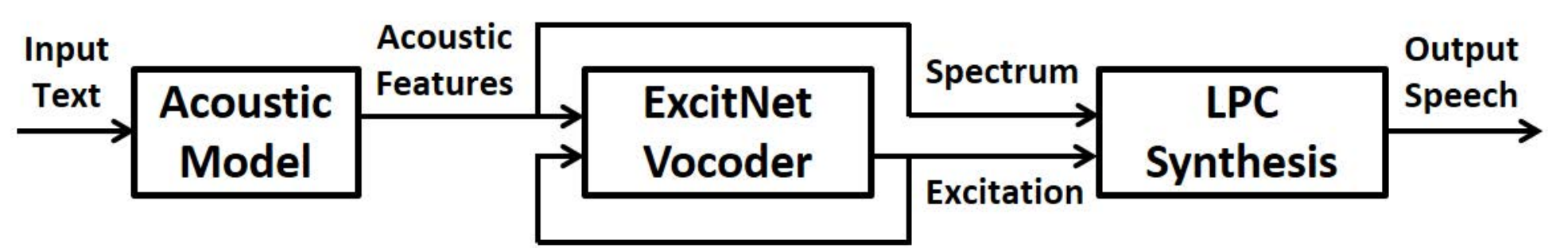,width=80mm}}
   		\centerline{(c)}  \medskip
    \vspace*{-6pt}  
   	\end{minipage}
   	\caption{Speech synthesis frameworks based on a conventional SPSS system with (a) an LPC vocoder, (b) a WaveNet vocoder, and (c) the proposed ExcitNet vocoder.}
   	\label{fig:welp_block} 
   	\end{figure}

\subsection{Speech Synthesis Using ExcitNet Vocoder}
\label{ssec:ch3-2}  
    Fig.~\ref{fig:welp_block}-(c) depicts a synthesis framework of the ExcitNet vocoder in which the architecture is combined from the conventional LPC vocoder presented in Fig.~\ref{fig:welp_block}-(a) and the WaveNet vocoder presented in Fig.~\ref{fig:welp_block}-(b). 
    To obtain the auxiliary features, we adopt our previous SPSS system based on the ITFTE vocoder \cite{song2017effective}.
    From the given text input, an \textit{acoustic model}\footnote{In this framework, the acoustic model consists of multiple feedforward and long short-term memory layers, trained to represent a nonlinear mapping function between linguistic input and acoustic output parameters. More detail about training the acoustic model is provided in the following section.} first estimates the acoustic parameters such as the LSF, F0, v\texttt{/}uv, gain, SEW, and REW, which are then used to compose the auxiliary features.
    By inputting these auxiliary features, the ExcitNet generates the time sequence of the excitation signal.
    Finally, the speech signal is reconstructed by passing the excitation signal through the LP synthesis filter formed by the generated LSFs.

\section{Experiments}
\label{sec:ch4}

	\begin{table*}[!t]
	\begin{center}
	\caption{Objective test results with respect to the different neural vocoders: In the SD systems, the amount of the training database is set to 1, 3, 5, and 7 hours. The systems that returned the smallest errors are in bold font.}
	\label{table:obj-v-spss}
	{\small        
	\begin{tabular}{>{\centering}m{.10\linewidth}|c||c|c|c|c|c||c|c|c|c|c}
	\Xhline{2\arrayrulewidth}
    \multicolumn{1}{c|}{\multirow{3}{*}{Speaker}}   & \multicolumn{1}{c||}{\multirow{3}{*}{System}} & \multicolumn{5}{c||}{LSD (dB)} & \multicolumn{5}{c}{F0 RMSE (Hz) } \\\cline{3-12}
    
    \multicolumn{1}{c|}{}   & \multicolumn{1}{c||}{}    & \multicolumn{4}{c|}{SD}   & \multicolumn{1}{c||}{\multirow{2}{*}{SI}} & \multicolumn{4}{c|}{SD}   & \multicolumn{1}{c}{\multirow{2}{*}{SI}}  \\\cline{3-6} \cline{8-11}
    
    \multicolumn{1}{c|}{} &\multicolumn{1}{c||}{} &\multicolumn{1}{c|}{1 h} &\multicolumn{1}{c|}{3 h} &\multicolumn{1}{c|}{5 h} &\multicolumn{1}{c|}{7 h} &\multicolumn{1}{c||}{}  &\multicolumn{1}{c|}{1 h} &\multicolumn{1}{c|}{3 h} &\multicolumn{1}{c|}{5 h} &\multicolumn{1}{c|}{7 h} &\multicolumn{1}{c}{}  \\ \hline\hline
	     	& WN        & 4.21	        & 4.19	        & 4.18	        & 4.13          & 4.18      & 32.61         & 31.69         & 31.56         & 31.49         & 32.30\\	
	KRF 	& WN-NS     & 4.15	        & 4.12	        & 4.07	        & 4.01          & 4.06      & 31.96         & 31.75         & 31.52         & 31.38         & 32.23\\	
	     	& ExcitNet  & \textbf{4.11}	& \textbf{4.09}	& \textbf{4.05}	& \textbf{3.99} & \textbf{4.04} &\textbf{31.44} &\textbf{31.43}	&\textbf{31.37}	&\textbf{31.29} & \textbf{31.88} 	\\	\hline 
	     	& WN        & 3.73	        & 3.72	        & 3.69	        & 3.67          & 3.70      & 12.60         & 12.39         & 12.05         & 12.05         & 13.96\\	
	KRM 	& WN-NS     & 3.54          & \textbf{3.46} & \textbf{3.41} & 3.41          & 3.46      &\textbf{12.32} & 12.16         & 11.97         & 11.97         & 13.34\\	
	     	& ExcitNet  & \textbf{3.53} & \textbf{3.46} & \textbf{3.41} & \textbf{3.40} & \textbf{3.45} & 12.72         &\textbf{12.10} &\textbf{11.93} &\textbf{11.93} & \textbf{12.96} \\	
			\Xhline{2\arrayrulewidth}
	\end{tabular}}
	\end{center}
    \vspace*{-8pt}
	\end{table*}

	\begin{table}[!t]   
	\begin{center}         
	\caption{Number of utterances in different sets.}  
	\label{table:numUtt}
	{\small        
	\begin{tabular}{>{\centering}m{.20\linewidth}||c|c|c}
	\Xhline{2\arrayrulewidth}
	Speaker			& Training  & Development 	& Test \\
			\hline \hline
	KRF 	& 3,826 (7 h)		& 270 (30 min) 			& 270 (30 min)	\\
			\hline
	KRM		& 2,294 (7 h)		& 160 (30 min)			& 160 (30 min)	\\
			\Xhline{2\arrayrulewidth}
	\end{tabular}}          
	\end{center}         
    \vspace*{-8pt}
	\end{table}

\subsection{Experimental Setup}
\label{ssec:ch4-1}

    To investigate the effectiveness of the proposed algorithm, we trained neural vocoding models using two different methods:
    \begin{itemize}
    \item SD: speaker-dependent training model
    \item SI: speaker-independent training model
    \end{itemize}
    Two phonetically and prosodically rich speech corpora were used to train the acoustic model and the SD-ExcitNet vocoder.
	Each corpus was recorded by professional Korean female (KRF) and Korean male (KRM) speakers.
	The speech signals were sampled at 24 kHz, and each sample was quantized by 16 bits.
	Table~\ref{table:numUtt} shows the number of utterances in each set.   
	To train the SI-ExcitNet \cite{hayashi2017investigation}, speech corpora recorded by five Korean female and five Korean male speakers were also used.
	In total, 6,422 (10 h) and 1,080 (1.7 h) utterances were used for training and development, respectively.
	The speech samples recorded by the same KRF and KRM speakers not included in the SI data set were used for testing.

    To compose the acoustic feature vectors, the spectral and excitation parameters were extracted using the ITFTE vocoder.
    The estimated 40-dimensional LP coefficients were converted into the LSFs for training.
    To prevent unnatural spectral peaks in the LP analysis/synthesis filter, each coefficient (${{a}_{i}, {i=1,...,40}}$) was multiplied by a linear prediction bandwidth expansion factor (${0.981^{i}}$) \cite{song2017effective}.
    In contrast, the 32-dimensional SEW and 4-dimensional REW coefficients were extracted for the excitation parameters.
    The F0, gain, and v\texttt{/}uv information were also extracted.
    The frame and shift lengths were set to 20 ms and 5 ms, respectively.
 
    In the acoustic modeling step, the output feature vectors consisted of all the acoustic parameters together with their time dynamics \cite{furui1986speaker}.
	The corresponding input feature vectors included 356-dimensional contextual information consisting of 330 binary features of categorical linguistic contexts and 26 numerical features of numerical linguistic contexts.
	Before training, both input and output features were normalized to have zero mean and unit variance.
	The hidden layers consisted of three feedforward layers (FFs) with 1,024 units and one unidirectional long short-term memory (LSTM) layer with 512 memory blocks.
    The FFs and LSTM were connected to the linguistic input layer and the acoustic output layer, respectively.
    The weights were initialized by \textit{Xavier} initializer \cite{xavier2010init} and trained using the \textit{backpropagation through time} (BPTT) algorithm with \textit{Adam} optimizer \cite{williams1990efficient, diederik2014adam}.
    The learning rate was set to 0.02 for the first 10 epochs, 0.01 for the next 10 epochs, and 0.005 for the remaining epochs.

    In the ExcitNet training step, all the acoustic parameters were used to compose the input auxiliary feature vectors, and they were duplicated from a frame to the samples to match the length of the excitation signals.
    Before training, they were normalized to have zero mean and unit variance.
    The corresponding excitation signal, obtained by passing the speech signal through the LP analysis filter, was normalized in a range from $-1.0$ to $1.0$ and quantized by 8-bit $\mu$-law compression.
    We used a one-hot vector to represent the resulting discrete symbol.
    The ExcitNet architecture had three convolutional blocks, each of which had ten dilated convolution layers with dilations of 1, 2, 4, 8, and so on up to 512. 
    The number of channels of dilated causal convolution and the 1$\times$1 convolution in the residual block were both set to 512.
    The number of 1$\times$1 convolution channels between the skip-connection and the softmax layer was set to 256.
    The learning rate was 0.0001, the batch size was 30,000 (1.25 sec), the weights were initialized using the Xavier initializer, and Adam optimizer was used.
	
    In the synthesis step, the mean vectors of all acoustic feature vectors were predicted by the acoustic model, and a speech parameter generation (SPG) algorithm was then applied to generate smooth trajectories for the acoustic parameters \cite{tokuda2000speech}.
    Because the acoustic model could not predict the variance used for the SPG algorithm, we used the pre-computed global variances of acoustic features from all training data \cite{ze2013statistical}.    
    By inputting these features, the ExcitNet vocoder generated a discrete symbol of the quantized excitation signal, and its dynamic was recovered via $\mu$-law expansion. 
    Finally, the speech signal was reconstructed by applying the LP synthesis filter to the generated excitation signal.
    To enhance spectral clarity, an LSF-sharpening filter was also applied to the generated spectral parameters \cite{song2017effective}.

	\begin{table}[!t]
	\begin{center}
	\caption{LSD (dB) test results measured in unvoiced and transition regions: The systems that returned the smallest errors are in bold font.}
	\vspace*{-4pt}
	\label{table:obj-uv}
	{\small        
	\begin{tabular}{>{\centering}m{.10\linewidth}|c||c|c|c|c|c}
	\Xhline{2\arrayrulewidth}
    \multicolumn{1}{c|}{\multirow{2}{*}{Speaker}}   & \multicolumn{1}{c||}{\multirow{2}{*}{System}} & \multicolumn{4}{c|}{SD} & \multicolumn{1}{c}{\multirow{2}{*}{SI}}    \\\cline{3-6}
    \multicolumn{1}{c|}{}                           & \multicolumn{1}{c||}{}                        & \multicolumn{1}{c|}{1 h}& \multicolumn{1}{c|}{3 h}& \multicolumn{1}{c|}{5 h}& \multicolumn{1}{c|}{7 h}   & \multicolumn{1}{c}{}  \\ \hline\hline
	     	& WN        & 4.19	        & 4.16	        & 4.15	        & 4.09          & 4.18	\\	
	KRF 	& WN-NS     & 4.26	        & 4.18	        & 4.12	        & 4.03          & 4.06	\\	
	     	& ExcitNet  & \textbf{4.15}	& \textbf{4.10}	& \textbf{4.06}	& \textbf{3.98} & \textbf{4.04} 	\\	\hline 
	     	& WN        & 3.95	        & 3.96	        & 3.92	        & 3.92          & 3.70	\\	
	KRM 	& WN-NS     & 4.41          & 3.95	        & 3.88          & 3.88          & 3.46	\\	
	     	& ExcitNet  & \textbf{3.91} & \textbf{3.83} & \textbf{3.76} & \textbf{3.76} & \textbf{3.45} \\	
			\Xhline{2\arrayrulewidth}
	\end{tabular}}	
	\end{center}   
    \vspace*{-8pt}
	\end{table}

\subsection{Objective Test Results}
\label{ssec:ch4-2}   

    To evaluate the performance of the proposed system, the results were compared to those of conventional systems based on a WaveNet vocoder (WN) \cite{tamamori2017speaker} and on a WaveNet vocoder with a noise-shaping method (WN-NS) \cite{tachibana2018investigation}.
    The WaveNet architectures and auxiliary features were the same with those of the proposed system, but the target outputs differed from each other.
    The target of the WN system was the distribution of the speech signal; that of the WN-NS system was the distribution of a noise-shaped residual signal.
    A time-invariant spectral filter in the latter system was obtained by averaging all spectra extracted from the training data \cite{fukada1992adaptive}.
    This filter was used to extract the residual signal before the training process, and its inverse filter was applied to reconstruct the speech signal in the synthesis step.

    In the test, distortions between the original speech and the synthesized speech were measured by log-spectral distance (LSD; dB) and F0 root mean square error (RMSE; Hz).
    Table~\ref{table:obj-v-spss} shows the LSD and F0 RMSE test results, with respect to the different neural vocoders.
    The findings can be analyzed as follows:
    (1) As the amount of training database increased in the SD systems, the overall estimation performances gradually improved for both the KRF and KRM speakers; 
	(2) In both the SD and SI systems, the vocoders with spectral filters (WN-NS and ExcitNet) achieved much more accurate speech reconstruction than the WN vocoder, which confirms that decoupling the formant component of the speech signal is beneficial to the modeling accuracy of the remaining signal;
	(3) Among the vocoders with spectral filters, ExcitNet's adaptive spectral filter helped to reconstruct a more accurate speech signal compared to the conventional system using a time-invariant filter (WN-NS).
	Since the average spectral filter in the WN-NS vocoder is biased towards voiced components, it is not optimal for unvoiced and transition regions which results in unsatisfactory results in those areas.
	This was clearly observed in the LSD results measured in the unvoiced and transition regions\footnote{As the F0 does not exists in the unvoiced and transition components, we only compared the LSD results in those regions.}, as shown in Table~\ref{table:obj-uv}.

    \begin{table}[!t]       
    \begin{center}         
    \caption{Subjective preference test results (\%) of synthesized speech for the KRF speaker: The systems that achieved significantly better preferences ($p~<~0.01$) are in bold.}
    \label{table:abtest-krf}
    {\small
    \begin{tabular}{>{\centering}m{.10\linewidth}||c|c|c|c|c}
    \Xhline{2\arrayrulewidth}
    KRF	    & WN	    & WN-NS	        & ExcitNet	    & Neutral   & p-value\\\hline \hline
    	    & 6.8		& \textbf{64.1}	& -		        & 29.1	    & $<10^{-30}$\\
    SD	    & 7.3		& -		        & \textbf{83.6}	& 9.1 	    & $<10^{-49}$\\
    	    & -			& 12.7  	    & \textbf{58.2}	& 29.1      & $<10^{-17}$\\  \hline	
    	    & 12.7		& \textbf{66.8}	& -         	& 20.5	    & $<10^{-22}$\\
    SI	    & 8.6		& -			    & \textbf{73.6}	& 27.7 	    & $<10^{-35}$\\
    	    & -			& 19.5		    & \textbf{38.6}	& 41.8	    & $<10^{-3}$\\
    \Xhline{2\arrayrulewidth}
    \end{tabular}}
    \end{center}  
    \vspace*{-8pt}
    \end{table} 	
    \begin{table}[!t]       
    \begin{center}         
    \caption{Subjective preference test results (\%) of synthesized speech for the KRM speaker: The systems that achieved significantly better preferences ($p~<~0.01$) are in bold.}
    \label{table:abtest-krm}
    {\small
    \begin{tabular}{>{\centering}m{.10\linewidth}||c|c|c|c|c}
    \Xhline{2\arrayrulewidth}
    KRM	    & WN	    & WN-NS	        & ExcitNet	    & Neutral   & p-value\\\hline \hline
    	    & 11.8  	& \textbf{60.5}	& -		        & 27.7	    & $<10^{-19}$\\
    SD	    & 17.3		& -		        & \textbf{77.7}	& 5.0 	    & $<10^{-24}$\\
    	    & -			& 16.4		    & \textbf{73.6}	& 10.0      & $<10^{-22}$\\  \hline	
    	    & 27.3		& \textbf{48.6}	& -         	& 24.1	    & $<10^{-3}$\\
    SI	    & 13.6		& -			    & \textbf{75.5}	& 10.9 	    & $<10^{-27}$\\
    	    & -			& 17.3		    & \textbf{63.6} & 19.1	    & $<10^{-15}$\\
    \Xhline{2\arrayrulewidth}
    \end{tabular}}
    \end{center}   
    \vspace*{-8pt}    
    \end{table} 	

\subsection{Subjective Test Results}
\label{ssec:ch4-3}

    To evaluate the perceptual quality of the proposed system, A-B preference and mean opinion score (MOS) listening tests were performed\footnote{Generated audio samples are available at the following url:\\ \url{https://sewplay.github.io/demos/excitnet}}.
    In the preference tests, 12 native Korean listeners were asked to rate the quality preference of the synthesized speech.
    In total, 20 utterances were randomly selected from the test set and were then synthesized using the three types of vocoder.
    Note that the input auxiliary condition features were obtained by the conventional SPSS system.
    Table~\ref{table:abtest-krf} and Table~\ref{table:abtest-krm} show the preference test results for the KRF and KRM speakers, respectively, and confirm that the perceptual quality of the speech synthesized by the ExcitNet vocoder is significantly better than those of the conventional WaveNet vocoders.

\begin{figure}[!t]
    \centerline{\epsfig{figure=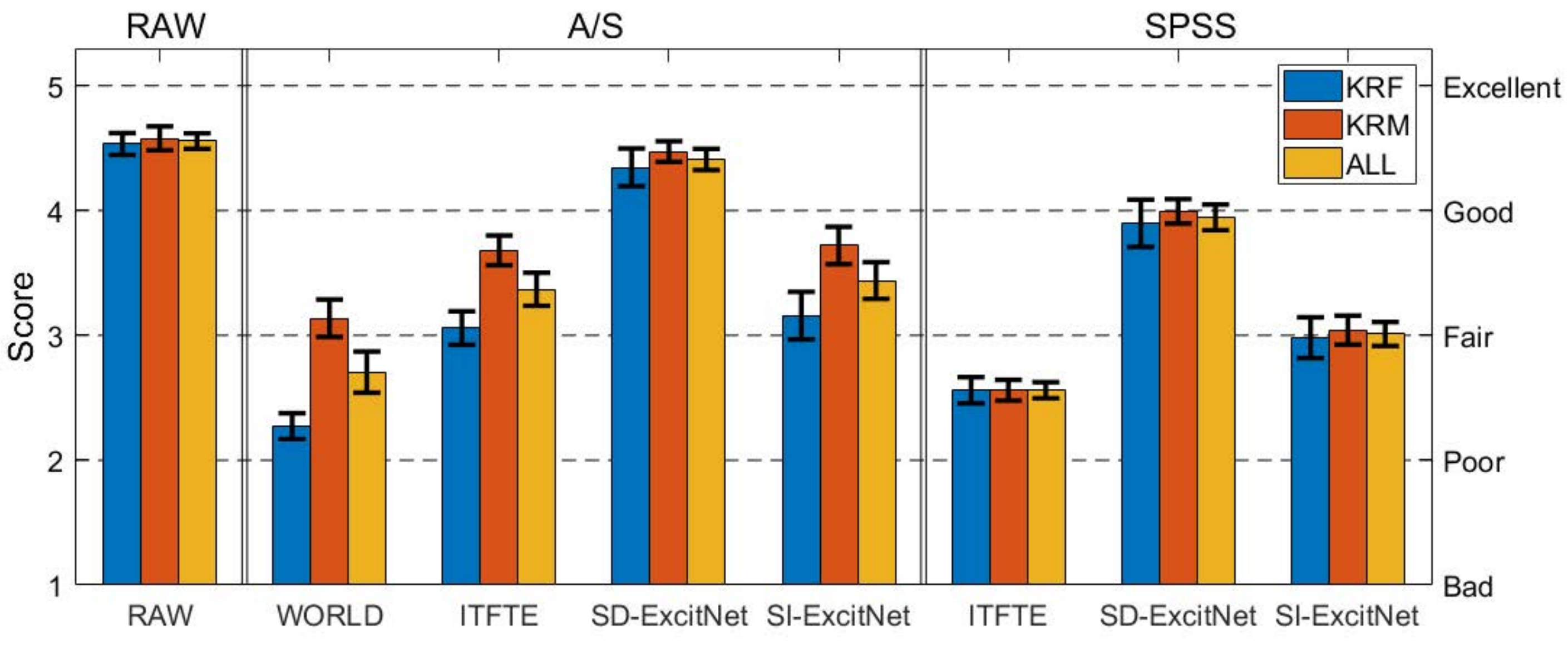,width=92mm}}
	\caption{Subjective MOS test results with 95\% confidence interval for previous and proposed systems. In the analysis/synthesis (A/S) and SPSS groups, acoustic features extracted from recorded speech and generated from the acoustic model, respectively, were used to compose the input auxiliary features.}
    \label{fig:mos}
    \end{figure}

    Setups for testing the MOS were the same as for the preference tests except that listeners were asked to make quality judgments of the synthesized speech using the following five possible responses: 1 = Bad; 2 = Poor; 3 = Fair; 4 = Good; and 5 = Excellent. 
    To verify vocoding performance, speech samples synthesized by conventional vocoders such as ITFTE and WORLD (D4C edition \cite{morise2016d4c}) were also included.
    The test results shown in Fig.~\ref{fig:mos} confirm the effectiveness of each system in several ways.
    First, the SI-ExcitNet performed similarly to the ITFTE vocoder but performed much better than the WORLD system in analysis/synthesis.
    Across all systems, the SD-ExcitNet provided the best perceptual quality (4.35 and 4.47 MOS for the KRF and KRM speakers, respectively).
    Next, owing to the difficulty of representing high-pitched female voices \cite{song2017perceptual}, the MOS results for the KRF speaker were worse than those for the KRM speaker in the SI vocoders (WORLD, ITFTE, and SI-ExcitNet).
    On the other hand, the results for the KRF speaker in the SD-ExcitNet were similar to those for the KRM speaker, which implies that modeling speaker-specific characteristics is necessary to represent high-pitched voices effectively.
    Lastly, in terms of SPSS, both the SD- and SI-ExcitNet vocoders provided much better perceptual quality than the parametric ITFTE vocoder.
    Although the acoustic model generated overly smoothed speech parameters, ExcitNet was able to alleviate the smoothing effect by directly estimating time-domain excitation signals.
    Consequently, the SPSS system with the proposed SD-ExcitNet vocoder achieved 3.78 and 3.85 MOS for the KRF and KRM speakers, respectively; the SI-ExcitNet vocoder achieved 2.91 and 2.89 MOS for the KRF and KRM speakers, respectively.

	\section{Conclusion}
	\label{sec:conclusion}
    This paper proposed the ExcitNet vocoder, built with hybrid architecture that effectively combined the merits of WaveNet and LPC vocoding structures.
    By decoupling the spectral formant structure from the speech signal, the proposed method significantly improved the modeling accuracy of the excitation signal using the WaveNet vocoder.
    The experimental results verified that the proposed ExcitNet system, trained either speaker-dependently or speaker-independently, performed significantly better than traditional LPC vocoders as well as similarly configured conventional WaveNet vocoders.
    Future research includes integrating the ExcitNet vocoder into speech synthesis systems that use an end-to-end approach.

\bibliographystyle{IEEEtran}
\bibliography{mybib}

\end{document}